\documentclass[aps,prd,twocolumn,groupedaddress,nofootinbib,]{revtex4-2}

\usepackage{booktabs}
\usepackage{amsmath}
\usepackage{array}
\usepackage{siunitx}
\usepackage{amssymb}
\usepackage{rotating}
\usepackage{graphicx}
\usepackage{xcolor}
\usepackage{lipsum}
\usepackage{ragged2e}
\usepackage{hyperref}
\usepackage{mathtools}
\usepackage{placeins}
\usepackage{orcidlink}
\usepackage{hyperref}
\def\ccr#1{{\color{black} #1}}
\def \be {\begin{equation}}
\def \ee {\end{equation}}
\def \ba {\begin{array}}
\def \ea {\end{array}}
\def \bea{\begin{eqnarray}}
\def \eea{\end{eqnarray}}

\begin{document}
\newcommand{\NCUa}{Department of physics, Nanchang University, Nanchang, 330031, China}
\newcommand{\NCUb}{Center for Relativistic Astrophysics and High Energy Physics, Nanchang University, Nanchang, 330031, China}
\newcommand{\SCUT}{School of Physics and Optoelectronics, South China University of Technology, Guangzhou 510641, China}

\title{Probing near-zone magnetic fields with extreme mass-ratio inspirals}

\author{Jin-Lu Hu}
\email{202420130571@mail.scut.edu.cn}
\affiliation{\SCUT}

\author{Xin-Dong Du}
\email{3537494784@qq.com}
\affiliation{\SCUT}

\author{Tieguang Zi\,\orcidlink{0000-0003-0046-2056}}
\email{zitieguang@ncu.edu.cn, corresponding author}
\affiliation{\NCUa}
\affiliation{\NCUb}

\author{Peng-Cheng Li\, \orcidlink{0000-0003-4977-2987}}
\email{pchli2021@scut.edu.cn, corresponding author}
\affiliation{\SCUT}

\begin{abstract}
We investigate whether weak near-zone magnetic fields can leave observable imprints on extreme-mass-ratio inspiral (EMRI) waveforms. The central massive black hole is modeled by the magnetized Schwarzschild, or Ernst, solution, and the secondary compact object is treated as a neutral point particle on equatorial circular geodesics. We compute the magnetic corrections to the circular-orbit quantities and the innermost stable circular orbit, and then evolve the inspiral using a hybrid, source-corrected Regge--Wheeler--Zerilli approximation, in which the Schwarzschild wave-propagation potentials are kept fixed while the source is evaluated on the magnetized orbit. For a fiducial system with \(M=10^6M_\odot\) and \(\mu=10M_\odot\), a field strength \(B\simeq 4\times10^{-5}M^{-1}\), corresponding to \(B_{\rm phys}\sim10^9\,{\rm G}\), produces a one-year dephasing of about \(1.3\) rad and \ccr{a LISA-noise-weighted mismatch which, after maximization over the relative arrival time and initial GW phase, remains slightly above the adopted distinguishability threshold at one year}. Our results suggest that EMRIs can in principle probe extremely strong near-zone magnetic fields, whereas ordinary magnetic environments around massive black holes are likely too weak to produce detectable effects within the present approximation.
\end{abstract}


\maketitle

\section{Introduction}
Gravitational-wave (GW) observations have opened a new window onto compact objects and strong-field gravity ~\cite{PhysRevLett.116.061102,LIGOScientific:2017vwq}. While current ground-based detectors are primarily sensitive to stellar-mass compact-binary coalescences, planned space-borne observatories such as LISA~\cite{LISA:2017pwj}, TianQin~\cite{10.1093/ptep/ptaa114}, and Taiji~\cite{Ruan:2018tsw} will probe the millihertz band, where extreme-mass-ratio inspirals (EMRIs) are among the most important sources \cite{Babak:2017tow}. An EMRI consists of a stellar-mass compact object orbiting a massive black hole (MBH) with a mass ratio ($\eta=\mu/M\sim10^{-7}-10^{-4}$). Because the secondary can complete a very large number of orbital cycles in the strong-field region before plunge, EMRI waveforms accumulate phase information over long timescales and are exceptionally sensitive to the central spacetime, possible deviations from general relativity \cite{Glampedakis:2005cf,Barack:2006pq,Cardoso:2019rvt,Zi:2021pdp,Zi:2023qfk,Zi:2023omh,Zi:2023geb,Datta:2024vll,Cardenas-Avendano:2024mqp,Zi:2024dpi,Zi:2024jla,Babichev:2024hjf,Kumar:2025jsi,Zare:2025aek,LaHaye:2025ley,Zhao:2025sck,Lu:2025xlp,Xia:2026aty,Long:2026dcb,Muguruza:2026hqn}, and astrophysical environmental effects \cite{Kocsis:2011dr,Barausse:2014tra,Pan:2021oob,LISA:2022kgy,Cardoso:2022whc,Speri:2022upm,Dai:2023cft,PhysRevD.107.023005,Zhao:2024bpp,Jiang:2024lwg,Mitra:2025tag,Vicente:2025gsg,Wang:2025rrj,Polcar:2025yto,Kejriwal:2025jao,Luo:2025ewp,Dyson:2025dlj,Haroon:2025rzx,Das:2025eiv,Rahman:2025mip,HegadeKR:2025dur,Zhao:2026yis,Zi:2026zpw,Azreg-Ainou:2026xcc}.

Among possible astrophysical environments, magnetic fields are particularly ubiquitous. Although isolated black holes are expected to be nearly electrically neutral~\cite{Gibbons:1975kk,10.1093/mnras/179.3.433}, large-scale magnetic fields can be supported by surrounding plasma and accretion flows. Observations of systems such as M87*~\cite{EventHorizonTelescope:2021bee,EventHorizonTelescope:2021srq}, Sagittarius A*~\cite{Eatough:2013nva}, and stellar-mass black-hole binaries provide evidence for magnetized environments near black holes~\cite{DelSanto:2012qt}. This raises a natural question: can such magnetic fields leave an observable imprint on EMRI gravitational waveforms?

In this paper, we adopt a hybrid, source-corrected approximation to estimate the leading imprint of a weak magnetic field on EMRI waveforms. The conservative orbital dynamics of the secondary compact object is treated exactly within the magnetized Schwarzschild, or Ernst, geometry~\cite{Ernst:1976mzr}: the particle follows equatorial circular geodesics of the magnetized background, and the corresponding orbital energy, angular momentum, azimuthal frequency, and the innermost stable circular orbit (ISCO) are computed from this spacetime. The dissipative sector, however, is modeled approximately. Instead of deriving the full gravitational-electromagnetic perturbation equations on the Ernst background, we use the standard Schwarzschild Regge-Wheeler-Zerilli (RWZ) formalism to compute the GW fluxes and reconstruct the waveform, with the source term evaluated along the magnetically corrected orbit. This approximation is motivated by two considerations. First, a fully self-consistent perturbation theory of the Ernst spacetime is technically challenging, since the background is non-spherical, non-asymptotically flat, and supported by an electromagnetic field, so gravitational and electromagnetic perturbations are expected to couple. Second, realistic astrophysical magnetic fields are likely to be localized in the near zone of the black hole and should not be interpreted as extending uniformly to spatial infinity. It is therefore natural to regard the Ernst solution as an effective near-zone description, matched at large radii to an asymptotically flat exterior where the usual GW flux and detector response can be defined. Our results should consequently be interpreted as an order-of-magnitude estimate of the magnetic-field-induced dephasing and mismatch, rather than as a complete waveform model for EMRIs in an exact magnetized black-hole spacetime.

Within this framework, we first analyze the magnetized background and the circular geodesic motion of the secondary compact object, including the fundamental orbital frequency, the conserved quantities, and the ISCO shift. We then evolve the orbit adiabatically using the source-corrected flux, construct the corresponding waveform, and quantify the accumulated dephasing and LISA-noise-weighted mismatch relative to the nonmagnetized Schwarzschild reference case.

The remainder of this paper is organized as follows. Section \ref{Ernst solution geometry and Magnetic field characteristics} presents the magnetized background, circular geodesic dynamics, ISCO shift, and source-corrected inspiral evolution. Section \ref{sec:results} gives the dephasing, waveform, and mismatch results. Section \ref{Conclusion} summarizes the main conclusions and discusses the limitations of our approximation.

Throughout this paper, we adopt the metric signature $(-,+,+,+)$ and use geometrized units with $G=c=1$.

\section{Method} \label{Ernst solution geometry and Magnetic field characteristics}
\subsection{Ernst Geometry and Magnetic-Field Characteristics}

A spacetime permeated by a uniform magnetic field can be obtained by solving the
Einstein--Maxwell equations. The corresponding electrovacuum solution is known as
the Melvin magnetic universe~\cite{Melvin:1963qx}. By applying a Harrison
transformation~\cite{Harrison:1968wue}, an asymptotically flat black-hole
solution can be embedded into the Melvin universe, leading to an exact
Einstein--Maxwell solution characterized by an external magnetic-field parameter.
For a Schwarzschild seed metric, this construction gives the magnetized
Schwarzschild, or Ernst, solution~\cite{Ernst:1976mzr}.\footnote{For theoretical and observational developments of magnetized black holes in recent years, see Refs.~\cite{Ernst:1976bsr,Liu:2020bag,Uktamov:2024zmj,Podolsky:2025tle,Wang:2025vsx,Li:2025rtf,
Siahaan:2026tuf,Zeng:2025olq,Tang:2025jop,Wang:2026czl,Liu:2025wwq,Zhang:2025ole,Yuan:2026knu,Vachher:2025jsq,Siahaan:2025ngu,Xamidov:2026kqs,Rehman:2026rzq,Zhang:2026myb,Hu:2026slp,Wan:2026lca}} 

In the present work, we consider the Schwarzschild black hole immersed in the
Melvin magnetic universe. In Schwarzschild-like coordinates, the line element can
be written as
\begin{equation}
ds^2=\Lambda^2\left[ -gdt^2+\frac{dr^2}{g}+r^2d\theta^2\right]
+\frac{r^2}{\Lambda^2}\sin^2\theta d\phi^2 ,
\label{eq:mag_sch_metric}
\end{equation}
where
\be
g=
1-\frac{2M}{r},\quad
\Lambda
=
1+\frac{1}{4}B^2 r^2\sin^2\theta.
\label{eq:Lambda_melvin}
\end{equation}
Here \(B\) denotes the strength of the magnetic field aligned with the symmetry
axis. The vector potential of the electromagnetic fields is given by
\be
A_\mu dx^\mu=\frac{2}{B}\left(1-\frac{1}{\Lambda} \right)d\phi.
\ee
Expanding in the magnetic-field parameter, the electromagnetic potential reduces at linear order in \(B\) to the test-field configuration of Wald on a Schwarzschild background~\cite{Wald:1974np}, while the metric deformation starts at order \(B^2\). Hence the Schwarzschild geometry is recovered in the limit \(B\to0\).

The coordinate location of the event horizon is determined by
\(g^{rr}=0\), and is therefore unchanged by the external magnetic field,
\begin{equation}
r_h=2M .
\end{equation}
Moreover, the horizon area remains \(A_h=16\pi M^2\), although the magnetic
field distorts the angular geometry away from spherical symmetry.

It is important to note that the metric in Eq.~\eqref{eq:mag_sch_metric} is not
asymptotically flat. Therefore, in this work
we regard this geometry as an effective near-zone description of the spacetime in
the region where the magnetic field generated by the surrounding plasma or
accretion disk is approximately uniform. At sufficiently large distances, the
near-zone magnetized geometry should be matched to an asymptotically flat
exterior. In the following analysis, we restrict our use of the Ernst metric to
the vicinity of the black hole, where this approximation is expected to capture
the leading influence of the magnetic field on the local geometry.

\ccr{More precisely, the present calculation does not construct a global
matched spacetime and therefore does not select a unique numerical matching
radius.  Denoting the characteristic magnetic length by $L_B=B^{-1}$, the
near-zone treatment requires $r_{\rm orb}/L_B=Br_{\rm orb}\ll1$.  We assume
that the physical magnetic field is localized and that an intermediate buffer
region exists in which
$r_{\rm orb}\ll R_{\rm match}\ll L_B$.  The value of $R_{\rm match}$ depends on
the global plasma and magnetic-field configuration and is not fixed by the
local Ernst model.  The weak-field waveform configurations considered below
satisfy this scale hierarchy by a wide margin.}

\subsection{Geodesic orbital dynamics}\label{Particle dynamics}

For a typical EMRI, the orbital period of the secondary is far less than 
the time-scale of inspiral into the MBH. Therefore, the orbital dynamics of the secondary can be treated as a sequence of geodesics around the MBH, with the orbital parameters updated adiabatically using the GW flux. Due to the existence of the Killing vectors $\partial_t$ and $\partial_\phi$ of the spacetime, the energy  and angular momentum  per unit mass of the geodesic orbit of the secondary are conserved quantities and are given by 
\be\label{EandLz}
E=-g_{tt}\dot{t},\quad L_z=g_{\phi\phi}\dot{\phi},
\ee
where the dot denotes the derivative with respect to  the proper time $\tau$. Moreover, the normalization of the four-velocity 
\be\label{velocitynorma}
g_{tt}\dot{t}^2+g_{rr}\dot{r}^2+g_{\theta\theta}\dot{\theta}^2+g_{\phi\phi}\dot{\phi}^2=-1,
\ee
provides the third conservation law.

Considering the secondary is moving on a circular orbit in the equatorial plane, one has $\theta=\pi/2$ and $r=\mathrm{const}$. Then from the above equation, one can obtain  the azimuthal fundamental frequency or the orbital angular velocity 
\begin{equation}
\Omega_{\phi} =\frac{d\phi}{dt}= \sqrt{-\frac{\partial_r g_{tt}}{\partial_r g_{\phi\phi}}}.
\end{equation}
In terms of the relation $\dot{\phi}=\Omega_{\phi}\dot{t}$, one can solve for  $\dot{t}$ from Eq. (\ref{velocitynorma}). Substituting the result into Eq. (\ref{EandLz}) gives
\begin{align}
E &= \frac{-g_{tt}}{\sqrt{-g_{tt} - g_{\phi\phi} \Omega_{\phi}^2}}, \\
L_z &= \frac{g_{\phi\phi} \Omega_{\phi}}{\sqrt{-g_{tt} - g_{\phi\phi}\Omega_{\phi}^2}}.
\end{align}
Substituting the explicit expressions of the metric of the spacetime Eq. (\ref{eq:mag_sch_metric})  gives \footnote{From now on  we will set $M=1$ for simplicity. In this case the magnetic field $B$ is dimensionless. The physical magnetic field in units of Gauss can be written as $B_{\mathrm{phys}}= 2.36\times 10^{19} \times B\times \left(\frac{M_\odot}{M}\right)$ G.}
\begin{equation}
\Omega_\phi =  \frac{1}{16} \sqrt{ -\frac{ \left(4 + B^2 r^2\right)^4 \left(4 + B^2 r^2 (-3 + 2r)\right) }{ r^3 \left(-4 + B^2 r^2\right) } },
\label{Omega}
\end{equation}
and
\begin{align}
E &=\frac{(-2 + r) (4 + B^2 r^2)}{4r \sqrt{ \frac{  \left( 12 + r \left( -4 + B^2 r (-5 + 3r) \right) \right) }{ r \left( -4 + B^2 r^2 \right) } }},\label{energy_r}
\\
L_z &=\frac{4 r \sqrt{  - \left(4 + B^2 r^2 (-3 + 2r)\right) } } { \left(4 + B^2 r^2\right) \sqrt{ \left(12 + r \left(-4 + B^2 r (-5 + 3r)\right)\right) } }.\label{angularmomentum_r}
\end{align}
 We can check that the above expressions reduce to the Schwarzschild case in the limit $B \to 0$.




\subsection{Innermost stable circular orbit shift}\label{Innermost stable circular orbit shift}
\begin{figure}
\includegraphics[width = 0.460\textwidth]{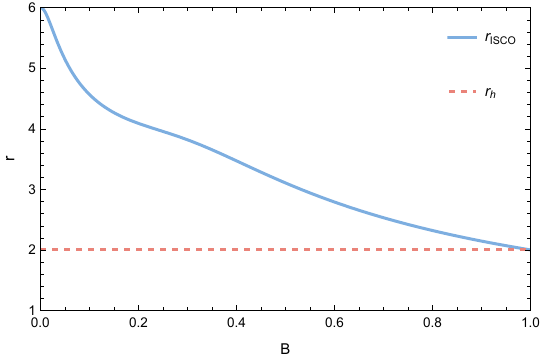}
\caption{The radius of ISCO  for various magnetic field parameter $B$. The blue line indicates the ISCO radius and the dashed line is the radius of event horizon.}
\label{Risco}
\end{figure}

The ISCO defines the smallest radius at which
a test particle can remain on a stable circular timelike geodesic. During
the adiabatic inspiral, the secondary compact object loses orbital energy and
angular momentum and gradually moves inward. Once the orbit reaches the ISCO,
the circular configuration becomes marginally stable; inside this radius the
particle can no longer maintain a stable circular orbit and subsequently plunges
toward the black hole. In the magnetized Schwarzschild geometry given by
Eq.~\eqref{eq:mag_sch_metric}, the external magnetic field breaks spherical symmetry but preserves stationarity, axisymmetry, and reflection symmetry with respect to the equatorial plane. We therefore restrict attention to equatorial circular geodesics. The ISCO can be determined either from the radial
effective potential or, equivalently, from the vanishing of the squared radial
epicyclic frequency. 

In terms of the orbital energy and angular momentum per unit mass $E$ and $L_z$, from the normalization of the four-velocity Eq. (\ref{velocitynorma}), we can derive the radial equation of the orbit in the equatorial plane
\be
\dot{r}^2 = V_{r}=-
g^{rr} \left( 1 + \dfrac{E^2}{g_{tt}} + \dfrac{L_z^2}{g_{\phi\phi}} \right). 
\ee
A circular orbit satisfies 
\begin{equation}
V_r(r)=0,
\qquad
\frac{dV_r}{dr}=0 ,\label{Vr}
\end{equation}
from which we can also derive the expressions of the  orbital energy and angular momentum per unit mass $E$ and $L_z$ as shown in Eqs.(\ref{energy_r}) and (\ref{angularmomentum_r}). Moreover, the marginal-stability condition defining the ISCO is given by
\begin{equation}
\frac{d^2V_r}{dr^2}=0.
\end{equation}
However, it turns out that the analytic form of the radius of the ISCO is very complicated. 

Alternatively, if the circular orbit is slightly perturbed, it oscillates with two characteristic epicyclic frequencies $\omega_r$ and $\omega_\theta$ in the radial and vertical direction, respectively. These frequencies can be obtained by perturbing the
equation (\ref{velocitynorma}) with
\begin{align}
r(t) &=r+\delta r(t),\\
\theta(t) &=\frac{\pi}{2} + \delta \theta(t).
\end{align}
Then after some calculations, the epicyclic frequencies are given by \cite{Ryan:1995wh}
\begin{align}
\omega_r &=\frac{1}{\sqrt{2g_{rr}}} \left[ g_{tt}^2 \partial_r^2 g^{tt} + \Omega_\phi^2 g_{\phi\phi}^2 \partial_r^2 g^{\phi\phi} \right]^{1/2} \Bigg|_{\theta=\pi/2},\\
\omega_\theta &= \frac{1}{\sqrt{2g_{\theta\theta}}} \left[ g_{tt}^2 \partial_\theta^2 g^{tt} + \Omega_\phi^2 g_{\phi\phi}^2 \partial_\theta^2 g^{\phi\phi} \right]^{1/2} \Bigg|_{\theta=\pi/2}.
\end{align}
Substituting the explicit expressions of the metric of the spacetime Eq. (\ref{eq:mag_sch_metric})  gives
\begin{align}
\omega_r &= \left[ \frac{ \mathcal{N}(r) }{ \mathcal{D}(r) } \right]^{1/2}, \\
\omega_\theta &= \sqrt{\frac{1}{r^3}},
\end{align}
where
\begin{align}
\mathcal{N}(r) &= 384 + r \bigl( -64 + B^2 r \bigl( -672 + r \bigl( 624 + r \bigl( -128 \notag \\
&\quad + B^2 \bigl( 200 + r \bigl( -204 + r \bigl( 48 + B^2 \bigl( -30 \notag \\
&\quad + (37 - 12r)r \bigr) \bigr) \bigr) \bigr) \bigr) \bigr) \bigr) \bigr), \\
\mathcal{D}(r) &= r^4 (B^2 r^2 - 4)(B^2 r^2 + 4)^2.
\end{align}
To determine the radius of the ISCO, one may impose
\begin{equation}
\omega_r^2=0 ,
\end{equation}
where \(\omega_r\) is the radial epicyclic frequency. Because the resulting
condition is a high-order algebraic equation, it is more convenient to solve it
numerically, following the standard method used for circular orbits in
non-Schwarzschild black-hole backgrounds~\cite{Doneva:2014uma,Abramowicz:2004tm}.

The marginal-stability condition of the orbit can be written as
\begin{align}
&12B^6 r^8
-37B^6 r^7
+\left(30B^6-48B^4\right)r^6
+204B^4 r^5
\nonumber\\
&\quad
+\left(128B^2-200B^4\right)r^4
-624B^2 r^3
+672B^2 r^2
\nonumber\\
&\quad
+64(r-6)=0 .
\label{eq:isco_polynomial}
\end{align}
The physical ISCO radius is selected as the real root of
Eq.~\eqref{eq:isco_polynomial} that lies outside the event horizon and is
continuously connected to the Schwarzschild value in the limit
\(B\rightarrow0\). In this limit, Eq.~\eqref{eq:isco_polynomial} reduces to
\begin{equation}
r_{\rm ISCO}=6,
\end{equation}
as expected. 

Equation~\eqref{eq:isco_polynomial} is solved numerically with
\texttt{Mathematica} for different values of \(B\). As shown in
Fig.~\ref{Risco}, the ISCO radius decreases as the magnetic-field strength
increases. For \(0<B<1\), the ISCO moves inward from the Schwarzschild value
toward the event horizon. At \(B=1\), the physical branch of the marginally
stable circular orbit reaches the horizon, \(r_{\rm ISCO}=r_h=2\). In realistic
astrophysical environments, however, the dimensionless magnetic field \(B\) is
expected to be much smaller than unity. Therefore, in the present work the
magnetic correction to the ISCO is treated as a small near-zone effect, and the
ISCO radius in this case can be simply expressed as
\begin{equation}
r_{\rm ISCO}\simeq 6 -864 B^2+\mathcal{O}(B^4).
\end{equation}

\subsection{Gravitational-wave flux and adiabatic evolution}\label{subsec:gw_flux_magnetic}

In this work we focus on the weak-magnetic-field regime, \(B\ll1\), and treat
the magnetized Schwarzschild geometry as a near-zone correction to the orbital
motion. A complete perturbative treatment of the Ernst spacetime would require
solving coupled gravitational and electromagnetic perturbation equations on a
non-asymptotically-flat and non-spherically-symmetric background. This is beyond
the scope of the present analysis. Instead, we adopt the 
RWZ equation with a magnetically corrected source term:
we use the standard Schwarzschild RWZ source formula and replace the circular-orbit frequency and constants of motion by their magnetically corrected values.

The RWZ master equation is written as
\begin{equation}
\left[
-\frac{\partial^2}{\partial t^2}
+\frac{\partial^2}{\partial r_*^2}
-V_l^{(P)}(r)
\right]
\Psi^{(P)}_{lm}(t,r)
=
S^{(P)}_{lm}(t,r;z^\mu_B),
\label{eq:rwz_master}
\end{equation}
where \(P=\mathrm{even},\mathrm{odd}\) denotes the parity sector,
\((l,m)\) are the spherical-harmonic indices, and
\[
r_*=r+2\ln\left(\frac{r}{2}-1\right),
\]
is the Schwarzschild tortoise coordinate. The source term
\(S^{(P)}_{lm}\) is determined by the worldline \(z^\mu_B\) of the secondary
compact object in the weakly magnetized background. The explicit expression of the source term is given in the Appendix \ref{Appdix} for convenience. The even-parity sector is
described by the Zerilli--Moncrief potential, whereas the odd-parity sector is
described by the Regge--Wheeler potential,
\begin{equation}
V_l^{(P)}(r)
=
\begin{cases}
V_l^{\rm ZM}(r), & P=\mathrm{even},\\
V_l^{\rm RW}(r), & P=\mathrm{odd}.
\end{cases}
\label{eq:rwz_potentials}
\end{equation}
Their explicit forms are standard and can be found, for example, in
Refs.~\cite{Martel:2005ir,Martel:2003jj}. For a particle moving on a circular
equatorial orbit, the reflection symmetry with respect to the equatorial plane
implies that the excited modes can be organized according to \(l+m\): modes with
\(l+m\) even contribute to the even-parity sector, while modes with \(l+m\) odd
contribute to the odd-parity sector.

The GW energy flux in each \((l,m)\) mode is computed from the inhomogeneous RWZ master solution evaluated at infinity and at the horizon. The energy fluxes can be written by summing over different modes
\begin{equation}
\dot E_{lm}^{\infty,H}
=
\frac{1}{64\pi}
\frac{(l+2)!}{(l-2)!}
\left(
\left|\dot{\Psi}^{\rm ZM,\infty,H}_{lm}\right|^2
+
4\left|\Psi^{\rm RW,\infty,H}_{lm}\right|^2
\right),
\label{eq:rwz_flux_lm}
\end{equation}
where \(\Psi^{\rm ZM}_{lm}\) and \(\Psi^{\rm RW}_{lm}\) denote the even- and
odd-parity inhomogeneous solutions, respectively. The superscripts
\(\infty\) and \(H\) indicate the outgoing solution at spatial infinity and the
ingoing solution at the horizon. The total flux is obtained by summing over
multipoles,
\begin{align}
\dot E^{\infty,H}
&=
\sum_{l=2}^{l_{\rm max}}
\sum_{m=-l}^{l}\dot E_{lm}^{\infty,H}.
\label{eq:flux_total_sum}
\end{align}
In the present calculation, we retain the leading quadrupolar
sector \(l=2\). Within this choice, the \(m=\pm1,\pm2\) modes dominate the radiated
power for circular motion. \ccr{As a truncation check, we also calculated the
$l=3$ and $l=4$ contributions.  The ratio
$(\dot E_{l=3}+\dot E_{l=4})/\dot E_{l=2}$ is typically about $10\%$ over much
of the interval $6\lesssim r\lesssim9.5$ and increases toward smaller radii,
reaching approximately $25\%$ near the ISCO.  We retain the $l=2$ flux for the
main parameter survey and quantify the resulting change in the accumulated
dephasing in Sec.~\ref{subsec:dephasing}.}

The above prescription should be understood as an approximation. 
The external magnetic field modifies the Ernst background curvature and hence, in a fully consistent calculation, also modifies the gravitational and electromagnetic perturbation equations and their effective potentials. 
Here we isolate the leading orbital effect of the magnetic field by keeping the Schwarzschild RWZ potentials fixed and incorporating the magnetic correction only through the source trajectory and the associated constants of motion. 
Therefore, the fluxes computed in Eqs.~\eqref{eq:rwz_flux_lm}--\eqref{eq:flux_total_sum} represent the source-induced correction to the Schwarzschild gravitational radiation, rather than the complete GW flux of the exact Ernst spacetime. 
The omitted radiative-sector corrections include the magnetic modification of the RWZ potentials, the near-zone-to-far-zone transfer function, and the coupling between gravitational and electromagnetic perturbations.

\ccr{A quantitative comparison of local weak-field scales can be obtained by
restoring $M$ and defining $x=r/M$ and $\beta=BM$.  On the equatorial plane,
\begin{align}
\frac{g_{tt}-g_{tt}^{\rm Sch}}{g_{tt}^{\rm Sch}}
&=\frac{1}{2}(\beta x)^2+\mathcal{O}(\beta^4x^4),\\
\frac{g_{\phi\phi}-g_{\phi\phi}^{\rm Sch}}{g_{\phi\phi}^{\rm Sch}}
&=-\frac{1}{2}(\beta x)^2+\mathcal{O}(\beta^4x^4),
\end{align}
whereas Eq.~\eqref{Omega} gives
\begin{equation}
\frac{\Omega_\phi-\Omega_\phi^{\rm Sch}}{\Omega_\phi^{\rm Sch}}
=\frac{x+1}{4}(\beta x)^2+\mathcal{O}(\beta^4x^4).
\label{eq:weak_field_scale_comparison}
\end{equation}
For $\beta=4\times10^{-5}$ and $6\lesssim x\lesssim9.2313$, the magnitude of
the fractional metric correction is approximately
$2.9\times10^{-8}$--$6.8\times10^{-8}$, while the fractional orbital-frequency
shift is approximately $1.0\times10^{-7}$--$3.5\times10^{-7}$, larger by a
factor of about $3.5$--$5.1$.  Since the RWZ source depends on the orbital
frequency and the conserved orbital quantities, this comparison motivates
isolating the orbital/source channel as the first contribution to study.  It
does not, however, prove that the potential correction is negligible: the
modified potentials and gravitational--electromagnetic coupling remain of the
same formal order in $B$ and could change or partially cancel the dephasing.
Accordingly, the results below should be interpreted as the
orbital/source-induced contribution and an order-of-magnitude estimate, rather
than as a complete $\mathcal{O}(B^2)$ prediction.}

\ccr{The radiation is not extracted at the non-asymptotically-flat infinity of
the exact Ernst solution.  Instead, the Schwarzschild RWZ outgoing solutions
and their asymptotically flat boundary conditions are used, thereby assigning
far-zone propagation and extraction to the assumed asymptotically flat
exterior.  Corrections associated with the matching region and the
near-zone-to-far-zone transfer function are not calculated here and form part
of the same systematic modeling uncertainty.}

For a typical EMRI, the orbital period is much shorter than the radiation-reaction
timescale. The inspiral can therefore be treated adiabatically. For a sequence of
quasi-circular orbits, the orbital radius evolves according to the flux-balance
relation
\begin{equation}
\frac{dr}{dt}
=
-\frac{\dot E^\infty+\dot E^H}{dE_B/dr},
\label{eq:rdot_balance}
\end{equation}
where \(E_B(r)\) is the specific orbital energy of a circular equatorial
geodesic in the weakly magnetized MBH. The minus sign reflects the loss of
orbital energy due to GW emission,
\begin{equation}
\frac{dE_B}{dt}=-\left(\dot E^\infty+\dot E^H\right),
\label{eq:energy_balance}
\end{equation}
The orbital frequency is then obtained from the magnetically corrected circular
geodesic relation,
\begin{equation}
\Omega_\phi^B=\Omega_\phi^B[r(t)],
\label{eq:Omega_B}
\end{equation}
For a circular orbit, each harmonic has frequency
\begin{equation}
\omega_m(t)=m\Omega_\phi^B[r(t)],
\label{eq:mode_frequency}
\end{equation}
Since the dominant GW contribution is the \((l,m)=(2,2)\) mode,
the leading GW phase satisfies
\begin{equation}
\frac{d\Phi_{\rm GW}}{dt}
=
2\Omega_\phi^B[r(t)],
\label{eq:gw_phase_frequency}
\end{equation}

To quantify the accumulated phase difference between an EMRI around the
magnetized black hole and the Schwarzschild reference case, we define the
dephasing as
\begin{equation}\label{eq:dephasing_magnetic}
\delta\Phi(t)
=
2\int_0^t
\left[
\Omega_\phi^B\!\left(r_B(t')\right)
-
\Omega_\phi^0\!\left(r_S(t')\right)
\right]dt',
\end{equation}
where \(\Omega_\phi^0\) denotes the orbital frequency of the corresponding
Schwarzschild inspiral, $r_B(t)$ and $r_S(t)$ denote the orbital evolution functions for the cases $B\neq0$ and $B=0$, respectively. The two inspirals are initialized at the same orbital radius and evolved for the same observation time. Since the magnetic field shifts the orbital frequency at fixed radius, the resulting dephasing contains both the initial frequency offset and the accumulated difference in the subsequent inspiral evolution. A nonzero
\(\delta\Phi\) therefore measures the cumulative GW phase shift
induced by the magnetic-field correction to the orbital motion. Following the commonly used dephasing criterion in EMRI studies~\cite{Datta:2019epe,Zi:2023geb}, we take an accumulated phase difference of order $\delta\Psi\simeq 1~{\rm radian}$ as a conservative indicator that two signals may become distinguishable by LISA for a source with signal-to-noise ratio $\rho\simeq20$.

\begin{figure}
\includegraphics{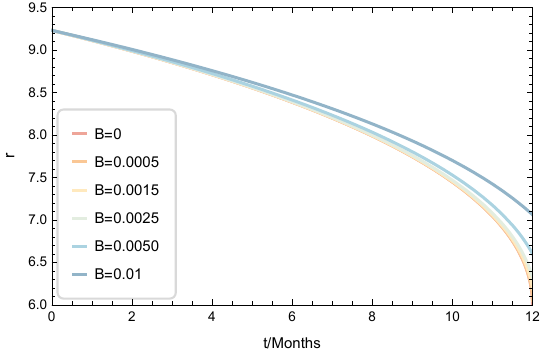}
\caption{
Evolution of the orbital radius \(r(t)\) over a one-year inspiral in the
magnetized Schwarzschild background for different magnetic-field parameters
\(B\). From top to bottom, the curves correspond to \(B=0.01\), \(0.0050\),
\(0.0025\), \(0.0015\), and \(0.0005\), respectively.
}\label{orbit evolution}
\end{figure}

\section{Results}\label{sec:results}
In this section, we present the effect of an external magnetic field on the
orbital evolution and gravitational waveforms of EMRIs. The inspiral is evolved
for one year within the perturbative framework described above, with the magnetic
field encoded by the parameter \(B\). Unless otherwise stated, the source
direction and the orientation of the MBH angular momentum are fixed to
$\theta_S=\pi/3,\phi_S=\pi/2,
\theta_K=\pi/4,\phi_K=\pi/4.$
The initial orbital separation \(r_0\) in the Schwarzschild case is chosen such that the system undergoes one year of adiabatic evolution before reaching the ISCO. All magnetized cases are initialized at the same radius 
and evolved for the same observation time. Since the ISCO radii in the magnetized cases are smaller than the Schwarzschild ISCO radius, the one-year magnetized inspirals terminate before reaching the Schwarzschild ISCO.  We consider a fiducial EMRI with component masses
$M=10^6M_\odot,\mu=10M_\odot,$
corresponding to a mass ratio \(\eta=\mu/M=10^{-5}\). The luminosity distance
\(D_L\) is treated as a free parameter, so that the signal-to-noise ratio can be
rescaled when needed.


\subsection{Frequency difference and dephasing}
\label{subsec:dephasing}

We first investigate how the magnetic field modifies the orbital evolution. We fix \(r_0=9.2313\), such that the Schwarzschild reference inspiral reaches $r_{\rm end}$ after one year. Using the adiabatic evolution equation in
Eq.~\eqref{eq:rdot_balance}, together with the GW phase relation
in Eq.~\eqref{eq:gw_phase_frequency}, we numerically evolve the orbital radius.
The resulting trajectories are shown in Fig.~\ref{orbit evolution}.

For weak magnetic fields, such as \(B=5\times10^{-4}\) and
\(B=1.5\times10^{-3}\), the orbital evolution almost coincides with the
Schwarzschild case. A visible deviation appears for stronger magnetic fields, for
example \(B=10^{-2}\). This behavior is expected because a larger magnetic field
modifies the circular-orbit constants of motion and the source term entering the
GW flux, thereby changing the accumulated radiation-reaction
effect during the inspiral.

\begin{figure}
\includegraphics[width=1\columnwidth]{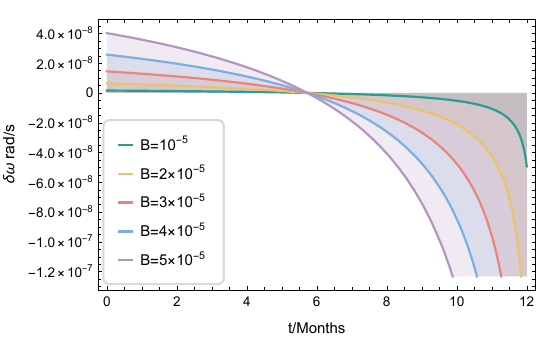}
\caption{
Difference in the azimuthal orbital frequency, \(\delta\omega\), between the
magnetized Schwarzschild inspiral and the vacuum Schwarzschild reference
inspiral over one year. Different curves correspond to different values of the
magnetic-field parameter \(B\). The frequency difference becomes more pronounced
as \(B\) increases.}\label{fig:omega:difference}
\end{figure}

We then compare the azimuthal orbital frequencies of EMRIs around the magnetized
MBH and the Schwarzschild MBH. The frequency difference as function of inspiral time is given by
\begin{equation}
\delta\omega(t)=2\left[  \Omega_\phi^B(r_B(t))-\Omega_\phi^0(r_S(t))\right],
\end{equation}
which is shown in Fig.~\ref{fig:omega:difference}
for several values of the magnetic-field parameter. The frequency difference is
not always positive during the one-year inspiral. It is first positive, reaches zero at approximately six months, and then becomes negative as the system approaches the late inspiral. As the magnetic-field strength increases, the
deviation from the Schwarzschild reference case becomes more pronounced.

\begin{figure}
\includegraphics[width=1\columnwidth]{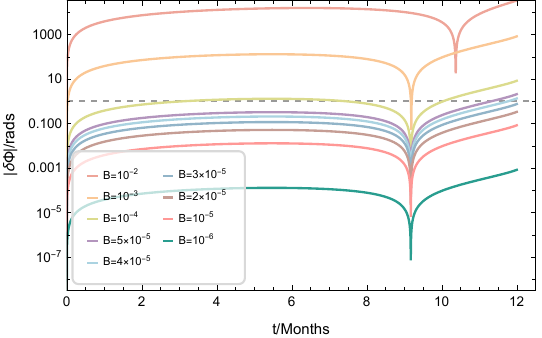}
\caption{
Accumulated dephasing \(\delta\Phi(t)\) over a one-year inspiral for different
magnetic-field parameters \(B\). From top to bottom, the curves correspond to
\(B=10^{-2}\), \(10^{-3}\), \(10^{-4}\), \(5\times10^{-5}\),
\(4\times10^{-5}\), \(3\times10^{-5}\), \(2\times10^{-5}\),
\(10^{-5}\), and \(10^{-6}\), respectively. The plotted quantity is the absolute value of the signed phase difference defined in the text, which allows us to use a logarithmic vertical axis. The sharp dips correspond to times when the signed accumulated dephasing passes close to zero. The gray dashed line marks the
reference value \(|\delta\Phi|=1\)  rad.}\label{dephasing}
\end{figure}

It is useful to decompose this quantity as
\begin{align}
\frac{\delta\omega(t)}{2}
=&\,
\left[
\Omega_\phi^B\!\left(r_B(t)\right)
-
\Omega_\phi^{0}\!\left(r_B(t)\right)
\right]
\nonumber\\
&+
\left[
\Omega_\phi^{0}\!\left(r_B(t)\right)
-
\Omega_\phi^{0}\!\left(r_{S}(t)\right)
\right].
\end{align}
The first term represents the direct magnetic correction to the orbital frequency at the same radius. In the weak-field regime considered here, this term is positive because the magnetic correction increases the azimuthal frequency at fixed radius. The second term represents the effect of the different orbital radii reached by the two inspirals at the same observation time. As shown in Fig.~\ref{orbit evolution}, the magnetized inspiral remains at a larger radius than the Schwarzschild reference inspiral, \(r_B(t)>r_{S}(t)\), after the evolution begins. Since the Schwarzschild circular-orbit frequency decreases with radius, the second term is negative. At early times, the two radii are still close and the positive direct magnetic correction dominates, giving \(\delta\omega>0\). At later times, the radial separation becomes larger and the negative radius-shift contribution can dominate, leading to \(\delta\omega<0\). This explains the zero crossing of the frequency difference in Fig.~\ref{fig:omega:difference} and the cancellation behavior of the accumulated phase difference shown in Fig.~\ref{dephasing}.

To quantify the cumulative impact of this frequency shift on the waveform phase,
we compute the dephasing defined in Eq.~\eqref{eq:dephasing_magnetic}. The
result is shown in Fig.~\ref{dephasing} as a function of observation time. As
expected, the overall magnitude, or envelope, of the accumulated absolute dephasing increases with the magnetic-field strength.
Using an accumulated phase shift of order one radian as a rough indicator of a
potentially observable effect, magnetic-field parameters around
\(B\sim10^{-5}-10^{-4}\) can already produce appreciable phase corrections for
the fiducial EMRI considered here. In particular, for \(B=4\times10^{-5}\), the
one-year accumulated dephasing reaches approximately \(1.331\) radian. For the illustrative
larger-field case \(B=10^{-2}\), the accumulated dephasing can reach
\(\mathcal{O}(10^3)\) radians.
\ccr{As a robustness check on the multipolar truncation, repeating the $B=4\times10^{-5}$ evolution with the summed $l=2,3,4$ flux gives a one-year dephasing of $1.277$ rad instead of $1.331$ rad.  The resulting change is approximately $4\%$, and the cancellation point is shifted only slightly, leaving the order of magnitude and the qualitative conclusions unchanged.}

The structure of the dephasing curves in Fig.~\ref{dephasing} can be understood from the frequency difference shown in Fig.~\ref{fig:omega:difference}. Since
\be
\frac{d\delta\Phi}{dt}=\delta\omega(t),
\ee
the signed dephasing is the accumulated signed area under the frequency-difference curve. At early times, the magnetic correction increases the orbital frequency at a fixed radius, giving a positive contribution to \(\delta\omega\). During the inspiral, however, the magnetized orbit remains at a larger radius than the Schwarzschild reference orbit at the same observation time, which tends to reduce the instantaneous orbital frequency. At late times this second effect can dominate, making \(\delta\omega\) negative. The positive and negative contributions to the accumulated phase can therefore partially cancel each other. The sharp dips in Fig.~\ref{dephasing} occur when the signed dephasing \(\delta\Phi\) passes close to zero, and hence appear as minima in the logarithmic plot of \(|\delta\Phi|\). These dips should therefore be interpreted as cancellation points of the accumulated phase difference, rather than as monotonic turning points of the physical dephasing.
For the weak-field cases \(B=10^{-6}\), \(10^{-5}\), and \(10^{-4}\), the
dips occur at nearly the same observation time, approximately
\(9.17\) months. For larger magnetic fields, the location of the dips
shifts slightly, occurring at about \(9.18\) months for \(B=10^{-3}\) and
\(10.37\) months for \(B=10^{-2}\). This shift reflects the stronger
modification of the orbital evolution when the magnetic field correction becomes
more significant.

\subsection{Waveform and mismatch}\label{sec:waveform_mismatch}
In this subsection, we present the gravitational waveform of an EMRI around a
weakly magnetized MBH and compare it with the corresponding Schwarzschild
reference waveform. To retain the relativistic information contained in the RWZ
formalism, we use the time-domain waveform reconstructed from the even- and
odd-parity master functions~\cite{Martel:2005ir,Martel:2003jj},
\begin{equation}
\begin{split}
h_+ - i h_\times
=
\frac{\mu}{2D_L}
\sum_{l=2}^{\infty}\sum_{m=-l}^{l}
\sqrt{\frac{(l+2)!}{(l-2)!}}
\biggl[
\Psi_{\rm ZM}^{lm}(t)
\\
-2i\int_{-\infty}^{t}\Psi_{\rm RW}^{lm}(t')dt'
\biggr]
{}_{-2}Y^{lm}(\theta,\phi),
\end{split}
\label{eq:rwz_waveform}
\end{equation}
where \(D_L\) is the luminosity distance, \({}_{-2}Y^{lm}\) are spin-weighted
spherical harmonics with spin weight \(s=-2\)~\cite{Goldberg:1966uu}, and
\(\Psi_{\rm ZM}^{lm}\) and \(\Psi_{\rm RW}^{lm}\) denote the Zerilli--Moncrief
and Regge--Wheeler master functions, respectively. In the present approximation,
the effect of the magnetic field enters the waveform through the magnetically
modified orbital trajectory and source term.

\begin{figure*}
\centering    
\includegraphics[width=0.45\textwidth]{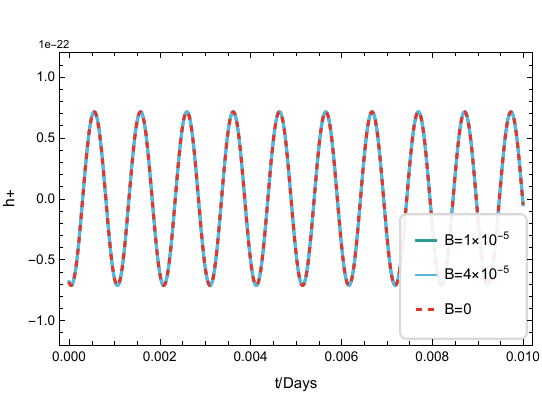}
\includegraphics[width=0.45\textwidth]{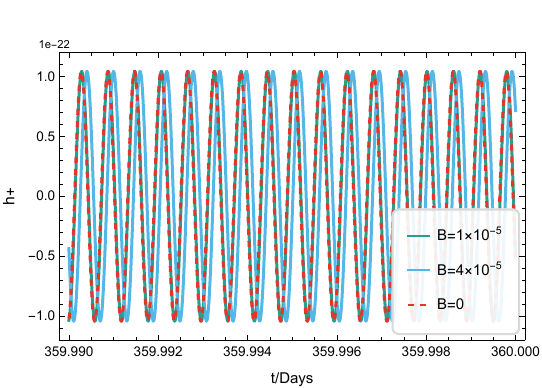}
\caption{
Comparison of the plus polarization \(h_+\) for EMRI waveforms generated by a
standard Schwarzschild black hole and by a Schwarzschild black hole in a
magnetic environment. The magnetized cases correspond to
\(B=1\times10^{-5}\) and \(B=4\times10^{-5}\). The left panel shows the
waveforms at the initial stage of the inspiral, while the right panel shows the
waveforms after one year of evolution.
}\label{waveform1}
\end{figure*}

Figure~\ref{waveform1} compares the waveforms for \(B=10^{-5}\),
\(B=4\times10^{-5}\), and the Schwarzschild case \(B=0\). At the beginning of
the inspiral, the three waveforms are almost indistinguishable. After one year of
evolution, the waveform with \(B=10^{-5}\) still remains close to the
Schwarzschild waveform, while the waveform with \(B=4\times10^{-5}\) develops a
visible phase offset. The accumulated phase difference is approximately
\(1.3\) rad, consistent with the dephasing shown in Fig.~\ref{dephasing}.

Under the low-frequency approximation, the response of a space-borne
interferometer can be written as
\begin{equation}
h(t)
=
\sum_{\alpha=I,II}
\frac{\sqrt{3}}{2}
\left[
F^+_\alpha(t)h_+(t)
+
F^\times_\alpha(t)h_\times(t)
\right],
\label{eq:lisa_response}
\end{equation}
where \(F^{+,\times}_{I,II}(t)\) are the time-dependent antenna pattern
functions. These functions depend on the sky location of the source and on the
orientation of the orbital angular momentum, specified by
\((\theta_S,\phi_S)\) and \((\theta_K,\phi_K)\), respectively
~\cite{Cutler:1997ta,Apostolatos:1994mx,Barack:2003fp}.

The dephasing discussed in the previous subsection provides a useful first
diagnostic of the magnetic-field effect. A more detector-oriented measure is the
mismatch between the waveform generated in the magnetized background and the
Schwarzschild reference waveform.
\ccr{Following the standard matched-filtering prescription~\cite{Owen:1995tm,Allen:2005fk}, we maximize over a relative arrival time $t_0$ and a common initial GW phase $\phi_0$.} The mismatch between two waveforms \(h_a\) and \(h_b\) is defined as
\begin{equation}
\ccr{\mathcal{M}=1-\mathcal{O}_{\rm max}(h_a,h_b)},
\label{eq:mismatch_def}
\end{equation}
where the \ccr{maximized} LISA-noise weighted overlap is
\begin{equation}
\ccr{\mathcal{O}_{\rm max}(h_a,h_b)}
=
\ccr{\max_{t_0,\phi_0}
\frac{\langle h_a|h_b(t_0,\phi_0)\rangle}
{\sqrt{\langle h_a|h_a\rangle\langle h_b|h_b\rangle}}} .
\label{eq:overlap_def}
\end{equation}
The noise-weighted inner product is given by
\begin{equation}
\langle h_a|h_b\rangle
=
2\int_0^\infty
\frac{
\tilde h_a^*(f)\tilde h_b(f)
+
\tilde h_a(f)\tilde h_b^*(f)
}
{S_n(f)}
df ,
\label{eq:inner_product}
\end{equation}
where \(\tilde h(f)\) is the Fourier transform of the time-domain detector
response, the star denotes complex conjugation, and \(S_n(f)\) is the one-sided
noise power spectral density of the detector~\cite{LISA:2017pwj,Robson:2018ifk}.

{\color{black}
For the positive-frequency part of the waveform, the two extrinsic shifts act as
\begin{equation}
\tilde h_b(f;t_0,\phi_0)
=e^{i(2\pi f t_0-\phi_0)}\tilde h_b(f).
\label{eq:time_phase_shift}
\end{equation}
It is useful to introduce the complex correlation
\begin{equation}
Z(t_0)=4\int_0^\infty
\frac{\tilde h_a^*(f)\tilde h_b(f)}{S_n(f)}
e^{i2\pi f t_0}\,df .
\label{eq:complex_correlation}
\end{equation}
At fixed $t_0$, maximizing the real inner product over the common phase gives
\begin{equation}
\max_{\phi_0}\langle h_a|h_b(t_0,\phi_0)\rangle=|Z(t_0)|,
\end{equation}
and hence
\begin{equation}
\mathcal{O}_{\rm max}
=\frac{\max_{t_0}|Z(t_0)|}
{\sqrt{\langle h_a|h_a\rangle\langle h_b|h_b\rangle}}.
\label{eq:maximized_overlap}
\end{equation}
We evaluate the time maximization with a complex inverse FFT.  For a sampling interval $\Delta t$ and a zero-padded length $N_{\rm FFT}$, with $\Delta f=(N_{\rm FFT}\Delta t)^{-1}$, we construct
\begin{equation}
Q_k=4\Delta f\,\frac{\tilde h_{a,k}^*\tilde h_{b,k}}{S_n(f_k)},
\qquad
Z_j=\sum_{k>0}Q_k e^{i2\pi jk/N_{\rm FFT}}.
\label{eq:discrete_complex_correlation}
\end{equation}
Each cumulative time segment is treated with the same window and is zero-padded to at least twice its original length, which prevents circular wrap-around in the correlation.  The one-year signal is sampled at 20 cumulative observation durations.  We maximize only over $t_0$ and the common GW phase $\phi_0$; no intrinsic source parameters are varied.  The resulting mismatch should therefore still be regarded as a distinguishability diagnostic rather than a full parameter-estimation result~\cite{Lindblom:2008cm}.
}

If two waveforms are identical, their overlap
is \(\mathcal{O}=1\) and the mismatch vanishes. If the waveforms are
uncorrelated, the overlap approaches zero, while perfectly anticorrelated
waveforms have \(\mathcal{O}=-1\). Following the usual waveform-accuracy
criterion, two signals are considered distinguishable when
\(\mathcal{M}\gtrsim \mathcal{D}/(2\rho^2)\)
~\cite{Lindblom:2008cm,Flanagan:1997kp}, where \(\mathcal{D}\) is the number of
intrinsic parameters and \(\rho\) is the signal-to-noise ratio. With $\mathcal{D}=8$ and 
\(\rho=20\) as a representative signal-to-noise ratio for LISA, the corresponding
distinguishability threshold is approximately \(\mathcal{M}\simeq10^{-2}\).

\begin{figure}
\includegraphics[width=1\columnwidth]{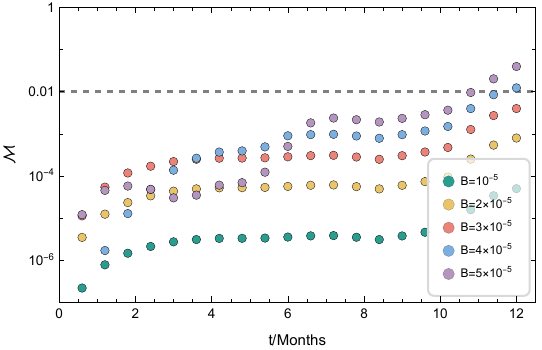}
\caption{
\ccr{Mismatch between EMRI waveforms with and without the external magnetic field after maximization over the relative arrival time $t_0$ and common initial GW phase $\phi_0$, shown as a function of cumulative observation time over a one-year inspiral.  The colored markers correspond to $B=1\times10^{-5}$, $2\times10^{-5}$, $3\times10^{-5}$, $4\times10^{-5}$, and $5\times10^{-5}$, as indicated in the legend.  The gray dashed line denotes the adopted distinguishability threshold $\mathcal{M}=0.01$.}
}\label{mismatch}
\end{figure}

\ccr{Figure~\ref{mismatch} shows the mismatch after maximization over $t_0$ and $\phi_0$.  The maximization substantially reduces the mismatch, as expected.  For $B\leq3\times10^{-5}$, it remains below the adopted threshold throughout the observation.  The $B=5\times10^{-5}$ curve exceeds the threshold only during the final part of the one-year evolution, while the $B=4\times10^{-5}$ curve lies slightly above the threshold at the 12-month point.  Thus the latter case is marginal, but it is not pushed below the threshold by maximizing over these two extrinsic parameters.}

\ccr{The maximized mismatch is not strictly monotonic with cumulative observation time.  This is not inconsistent with its definition: changing the duration changes the normalized noise-weighted inner product and also leads to a new optimal time and phase alignment.  We also observe a nonmonotonic ordering with $B$ during approximately the first six months, where the absolute mismatches are small.  This early-time ordering is unchanged by our numerical convergence tests, including variations of the sampling interval and zero-padding length.  We additionally checked the leading weak-field expectation $\mathcal{M}\propto B^4$; the $B=4\times10^{-5}$ and $5\times10^{-5}$ cases do not follow a simple $B^4$ rescaling.  Such proportionality is an asymptotic leading-order expectation rather than an exact identity for a finite-duration, normalized overlap that is re-maximized for each $B$.  We therefore do not assign physical significance to the detailed early-time ordering or use it to infer the critical field strength.  Our robust conclusion is restricted to the late-time values and, in particular, to the marginal one-year result for $B=4\times10^{-5}$.}

\section{Conclusions}\label{Conclusion}

In this work, we have investigated the impact of an external magnetic field on
the orbital dynamics and GW signatures of EMRIs around a
MBH. The background spacetime was modeled by the magnetized
Schwarzschild, or Ernst, solution, which describes a Schwarzschild black hole
embedded in the Melvin magnetic universe. Since this spacetime is not
asymptotically flat, we have interpreted it as an effective near-zone
description of the magnetized environment around the MBH.

We first studied equatorial circular geodesics of a neutral compact object in
the magnetized background. The external magnetic field modifies the orbital
energy, angular momentum, and azimuthal frequency, and shifts the location of
the innermost stable circular orbit. We confirmed that the ISCO moves inward as
the magnetic-field parameter increases, while the Schwarzschild result
\(r_{\rm ISCO}=6\) is recovered in the limit \(B\rightarrow0\).

We then estimated the GW emission and adiabatic inspiral within
a source-corrected RWZ approximation. In this treatment, the
Schwarzschild RWZ potentials are kept fixed, while the magnetic-field correction
is incorporated through the modified orbital trajectory and source terms. This
approximation isolates the leading orbital effect of the magnetic field on the
waveform phase, but it does not represent a complete perturbation theory of the
Ernst spacetime. A fully consistent calculation would require solving the
coupled gravitational and electromagnetic perturbation equations on the
magnetized background.

\ccr{The weak-field scale comparison in Sec.~\ref{subsec:gw_flux_magnetic}
shows that the fractional orbital-frequency shift is a few times larger than
the local fractional metric deformation over the fiducial one-year orbit.  This
supports isolating the orbital/source channel as a first estimate, but it does
not establish that the omitted potential correction is negligible.  The latter
is of the same formal order in $B$ and may reinforce or partially cancel the
reported source-induced dephasing.}

For the fiducial system considered in this paper, with
\(M=10^6M_\odot\) and \(\mu=10M_\odot\), we evolved one-year quasi-circular
inspirals and compared the resulting waveforms with the Schwarzschild reference
case. The accumulated dephasing grows with the magnetic-field strength. In
particular, a magnetic parameter \(B=4\times10^{-5}\) produces a one-year
dephasing of approximately \(1.331\) rad, while stronger magnetic fields can
lead to much larger phase differences. \ccr{Including the $l=3$ and $l=4$ fluxes in a dedicated $B=4\times10^{-5}$ truncation check changes the one-year dephasing from $1.331$ rad to $1.277$ rad, or by approximately $4\%$, supporting the robustness of its order of magnitude.  The waveform mismatch gives a more detector-oriented measure of distinguishability.  After maximizing the overlap over the relative arrival time and common initial GW phase, the mismatch remains below the adopted LISA threshold for $B\leq3\times10^{-5}$.  The $B=5\times10^{-5}$ case crosses the threshold only late in the observation, and the $B=4\times10^{-5}$ case remains slightly above it at one year.  The latter result has little margin and should therefore be interpreted as a model-dependent, order-of-magnitude estimate rather than a sharp minimum detectable field strength.}

Converting the dimensionless magnetic-field parameter to physical units, our
results indicate that magnetic fields of order \(10^9\,{\rm G}\) around a
\(10^6M_\odot\) MBH may produce observable EMRI waveform
modifications. This field strength is much larger than typical estimates for many massive black-hole environments, so the result should also be interpreted as an order-of-magnitude upper benchmark for when magnetic-field corrections become relevant to EMRI phasing.

\ccr{For comparison with other EMRI environmental signatures, thin-disk
migration models can produce phase shifts of approximately $10$--$10^3$ rad per
year, depending strongly on the disk model and on whether a gap is present
\cite{Kocsis:2011dr}, while rotating dark-matter spikes can also produce
potentially detectable dephasings and mismatches with LISA
\cite{Mitra:2025tag}.  These comparisons are necessarily qualitative because the
environmental profiles, waveform prescriptions, and source parameters differ.
The approximately one-radian magnetic dephasing found here lies within the
broad range of potentially measurable environmental phase corrections, but it
requires an unusually strong magnetic field.  Even for active galactic nuclei,
estimated horizon fields have a mean of order $10^4\,{\rm G}$, with individual
objects reaching approximately $10^5$--$10^6\,{\rm G}$
\cite{Piotrovich:2020}; these values remain about three to five orders of
magnitude below the $\sim10^9\,{\rm G}$ scale inferred in our fiducial model.}

Future work should extend this
analysis by deriving the full perturbation equations on the Ernst background,
including the coupled gravitational--electromagnetic sector, and by performing
Bayesian parameter estimation with more realistic EMRI waveforms and detector
responses.

\section{ACKNOWLEDGMENTS}
\label{ACKNOWLEDGMENTS}
We are grateful to Jian-dong Zhang and An Gong for  helpful discussions. This work makes use of \textit{Black Hole Perturbation Toolkit} and was in part supported by NSFC Grant No.12205104. T.Zi is funded by the National Natural Science Foundation of China with Grant No. 12347140 and No. 12405059.

\appendix
\renewcommand{\theequation}{A.\arabic{equation}}
\section{PERTURBATION DETAILS}
\label{Appdix}

The source terms in the RWZ equation are given by \cite{Martel:2003jj,Martel:2005ir}
\begin{align}
S_{ZM} &= \frac{1}{(\lambda + 1)\mathcal{A}} \left\{ r^2 f \left( f^2 \frac{\partial}{\partial r} Q^{tt} - \frac{\partial}{\partial r} Q^{rr} \right) \right. \notag \\
&\quad \left. + r(\mathcal{A} - f) Q^{rr} + r f^2 Q^b \right. \notag \\
&\quad \left. - \frac{f^2}{r\mathcal{A}} \left[ \lambda(\lambda - 1)r^2 + (4\lambda - 9)r + 15 \right] Q^{tt} \right\} \notag \\
&\quad + \frac{2f}{\mathcal{A}} Q^r - \frac{f}{r} Q^{\#}, \\
S_{RW} &= \frac{f}{r} \left[ \frac{2}{r} \left( 1 - \frac{3}{r} \right) P - f \frac{\partial}{\partial r} P + P^r \right],
\end{align}
where $f=1-2/r$, $\lambda=(l+2)(l-1)/2$ and $\mathcal{A}=\lambda+3/r$. In the source-corrected prescription used in this work, these standard Schwarzschild RWZ source expressions are evaluated with the magnetically corrected orbital frequency and constants of motion.

These equations are obtained by projecting the energy-momentum tensor of a point particle $T^{\mu\nu}=\mu \int  \frac{dz^\mu}{d\tau}\frac{dz^\nu}{d\tau} \frac{\delta^4(x^\alpha - z^\alpha(\tau))}{\sqrt{-g}} \, d\tau $
onto spherical harmonic functions $Y^{lm}$, where $g$ is the determinant of the metric, and $z^\alpha(\tau)$ is the coordinate of a particle moving along a geodesic.
\begin{align}
Q^{ab} &= 8\pi \int T^{ab} Y^{lm*} \, d\Omega, \\
Q^a &= \frac{16\pi r^2}{l(l+1)} \int T^{aA} Z^{lm*} \, d\Omega,\\
Q^b &= 8\pi r^2 \int T^{AB} U^{lm*} \, d\Omega, \quad \\
Q^\# &= \frac{32\pi r^4}{(l-1)l(l+1)(l+2)} \int T^{AB} V^{lm*} \, d\Omega,
\end{align}
\begin{align}
P^a &= \frac{16\pi r^2}{l(l+1)} \int T^{aA} X_A^{lm*} \, d\Omega, \quad \\
P &= \frac{16\pi r^4}{(l-1)l(l+1)(l+2)} \int T^{AB} W_{AB}^{lm*} \, d\Omega,
\end{align}
where lower-case roman indices run over $t$ and $r$, and capital roman indices run over the angular coordinates $\theta$ and $\phi$, and $d\Omega=\sin{\theta}d{\theta}d{\phi}$. In terms of these quantities, the vectorial and tensorial spherical harmonics are defined as
\begin{align}
Z_A^{lm} &= Y_{|A}^{lm}, \quad \\
X_A^{lm} &= \varepsilon_A^{\ B} Y_{|B}^{lm}, \quad \\
U_{AB}^{lm} &= \Omega_{AB} Y^{lm}, \quad \\
V_{AB}^{lm} &= Y_{|AB}^{lm} + \frac{l(l+1)}{2} \Omega_{AB} Y^{lm}, \quad \\
W_{AB}^{lm} &= X_{(A|B)}^{lm},
\end{align}
where \(\Omega_{AB}=\mathrm{diag}(1,\sin^2\theta)\). A vertical bar denotes the covariant derivative compatible with $\Omega_{AB}$ and $\varepsilon_{AB}$ is the Levi-Civita tensor on the unit two-sphere.

\bibliographystyle{apsrev4-2}   
\providecommand{\noopsort}[1]{}\providecommand{\singleletter}[1]{#1}%

\end{document}